\documentclass[aip,apl,reprint]{revtex4-1}

\usepackage[T1]{fontenc}
\usepackage{textcomp}
\usepackage[sort&compress]{natbib}

\usepackage{amsmath}
\usepackage{txfonts}

\usepackage{graphicx}

\begin{document}

\title{Stoichiometry control of the electronic properties of the LaAlO$_{\text{3}}$/SrTiO$_{\text{3}}$ heterointerface}

\author{H. K. Sato}
\email{1229893981@mail.ecc.u-tokyo.ac.jp}
\affiliation{Stanford Institute for Materials and Energy Sciences, SLAC National Accelerator Laboratory, Menlo Park, California 94025, USA}
\affiliation{Department of Advanced Materials Science, University of Tokyo, Kashiwa, Chiba 277-8561, Japan}

\author{C. Bell}
\affiliation{Stanford Institute for Materials and Energy Sciences, SLAC National Accelerator Laboratory, Menlo Park, California 94025, USA}

\author{Y. Hikita}
\affiliation{Stanford Institute for Materials and Energy Sciences, SLAC National Accelerator Laboratory, Menlo Park, California 94025, USA}

\author{H. Y. Hwang}
\affiliation{Stanford Institute for Materials and Energy Sciences, SLAC National Accelerator Laboratory, Menlo Park, California 94025, USA}
\affiliation{Geballe Laboratory for Advanced Materials, Stanford University, Stanford, California 94305-4045, USA}

\date{\today}

\begin{abstract}
We investigate the effect of the laser parameters of pulsed laser deposition on the film stoichiometry and electronic properties of LaAlO$_{\text{3}}$/SrTiO$_{\text{3}}$ (001) heterostructures.
The La/Al ratio in the LaAlO$_{\text{3}}$ films was varied widely from 0.88 to 1.15, and was found to have a strong effect on the interface conductivity.
In particular, the carrier density is modulated over more than two orders of magnitude.
The film lattice expansion, caused by cation vacancies, is found to be the important functional parameter.
These results can be understood to arise from the variations in the electrostatic boundary conditions, and their resolution, with stoichiometry.
\end{abstract}
\maketitle

The discovery of a conducting layer formed at the interface between the wide band gap perovskite insulators LaAlO$_{3}$ (LAO) and SrTiO$_{3}$ (STO)\cite{Nature-427-423} has led to extensive research on this system.
Several scenarios have been proposed to explain the origin of the conductivity: an electronic reconstruction driven by the polar discontinuity between the $(001)$ interfaces of polar LAO and nonpolar STO,\cite{NatureMater-5-204} the formation of oxygen-deficient SrTiO$_{3-\delta}$ during growth,\cite{PhysRevB-75-121404,PhysRevLett-98-196802,PhysRevLett-98-216803} or atomic diffusion near the interface to create conducting La$_{x}$Sr$_{1-x}$TiO$_{3}$ layers.\cite{PhysRevLett-99-155502,SurfSciRep-65-317}
All are possible contributing mechanisms which are not mutually exclusive, and can interact in a highly complex way.\cite{arXiv:1105.5779}

In order to disentangle these effects, a central challenge experimentally is to understand the sensitivity of the properties of this system to all of the standard control parameters in thin film growth.
These include the oxygen partial pressure,\cite{Nature-427-423,NatureMater-6-493} the post-annealing treatment,\cite{NatureMater-7-621,EurophysLett-91-17004} the growth temperature\cite{PhysRevLett-105-236802} and the LAO thickness.\cite{Science-313-1942,ApplPhysLett-94-222111,EurophysLett-91-17004}
The latter effect, in particular the existence of a critical thickness of 3--4 unit cells (uc) required for conductivity,\cite{Science-313-1942} can be interpreted to support the polar discontinuity picture.
However, it is clear that there are other, larger, characteristic thicknesses which also influence the conductivity,\cite{,ApplPhysLett-94-222111, EurophysLett-91-17004} indicating the need of more detailed considerations.
The strong asymmetry between the LaO/TiO$_{2}$ (often called $n$-type in the literature) and AlO$_{2}$/SrO ($p$-type) interfaces, in particular the insulating nature of the latter\cite{Nature-427-423,JpnJApplPhys-43-L1032} as well as the structural differences,\cite{NatureMater-5-204,PhysRevLett-107-036104} not only emphasize the importance of the atomic scale details, but also imply that oxygen vacancies and intermixing are not the sole cause of the conductivity.

In the case of pulsed laser deposition (PLD), the character of the ablating laser itself is well known in oxides to be crucial to obtain precisely controlled film stoichiometry.\cite{JApplPhys-86-6528,JApplPhys-103-103703,AdvMater-20-2528}
Although most LAO/STO samples are grown by PLD, the laser parameters are often not explicitly described in many publications, and the study of their influence on the properties of this system is currently limited.\cite{PhysRevB-83-085408,JPhysCondensMatter-23-305002}
Indeed, \emph{a priori} the influence of stoichiometry variations of the LAO on the conductivity in the underlying STO is rather indirect, compared to the typical optimization of electronic properties of conducting thin films by directly tuning their own cation ratio.
However, it is still quite possible that the above scenarios used to explain the interface conductivity can be significantly impacted by a change in the LAO stoichiometry.
In this letter, we report electronic properties of various LAO/STO heterostructures with LAO grown using different laser fluences, demonstrating a dramatic influence on the interface conductivity.

The samples were fabricated by PLD using a KrF excimer laser, in an on-axis geometry with a substrate--target distance of 55 mm.
The laser repetition rate was fixed at 2 Hz, unless otherwise indicated.
For the x-ray diffraction (XRD) and transport measurements, each LAO film was grown on a $5 \times 5 \text{ mm}^{2}$ STO (001) substrate with a TiO$_{\text{2}}$-terminated surface.\cite{Science-266-1540,ApplPhysLett-73-2920}
The variable growth parameters were: the laser spot size $A$, the laser energy $E$, and the laser fluence $f = E/A$.
The laser beam was imaged to a rectangular spot on the single crystal LAO target using a four lens afocal zoom stage.
Before growth, the substrates were preannealed at $950 \text{ \textcelsius}$ in an oxygen partial pressure of $5 \times 10 ^{-6} \text{ Torr}$ for 30 mins.
Following this anneal, the substrate was cooled to $800 \text{ \textcelsius}$ and the oxygen partial pressure was increased to $1 \times 10 ^{-5} \text{ Torr}$, the film growth conditions.
The LAO film thickness was 25 uc, as monitored using \emph{in situ} reflection high-energy electron diffraction.
After growth, the samples were cooled to room temperature in an oxygen partial pressure of $300 \text{ Torr}$, with a one hour pause at $600 \text{ \textcelsius}$.
These growth details (other than the laser parameters) are the same as used previously.\cite{ApplPhysLett-94-222111}

The out-of-plane lattice constant of the LAO films ($c_{\text{LAO}}$) was evaluated using the (002) peak of the XRD $\theta$--$2 \theta$ scans.\cite{LAOindex}
For transport measurements, the samples were electrically contacted using an ultrasonic wirebonder with aluminum wire, in a Hall bar configuration with voltage contacts $\sim$1 mm apart.
The cation stoichiometry of the LAO films was measured by inductively coupled plasma (ICP) spectrometry, using $\sim$400 nm thick amorphous LAO films grown on $10 \times 10 \text{ mm}^{2}$ B-doped Si substrates at room temperature.
For these samples, in order to obtain the necessary thick films in a practical time, the repetition rate of the laser was increased to 10 Hz, while the oxygen pressure was the same as used for the epitaxial growth.

\begin{table}
\centering
\caption{Cation ratio from ICP and out-of-plane lattice constant of LaAlO$_{3}$ films grown at different laser conditions.}
\begin{tabular}{cccc}
\hline \hline
Fluence & Spot area & La/Al & $c_{\text{LAO}}$ \\
$f $ (J/cm$^{2}$) & $A$ (mm$^{2}$) & atomic ratio & ({\AA}) \\ \hline
0.7 & 3.0 & 0.984 & 3.726 \\
0.7 & 4.3 & 0.969 & 3.735 \\
0.7 & 5.6 & 0.996 & 3.737 \\
0.9 & 3.0 & 0.965 & 3.733 \\
1.1 & 3.0 & 0.957 & 3.739 \\
1.1 & 3.5 & 0.908 & 3.749 \\
1.3 & 2.5 & 0.883 & 3.753 \\
1.3 & 3.0 & 0.946 & 3.742 \\
1.6 & 2.5 & 1.155 & 3.763 \\
1.6 & 3.0 & 0.992 & 3.748 \\
1.9 & 2.5 & 1.037 & 3.758 \\
\hline \hline
\end{tabular}
\label{tb:tb1}
\end{table}

\begin{figure}[b]
\centering
\includegraphics{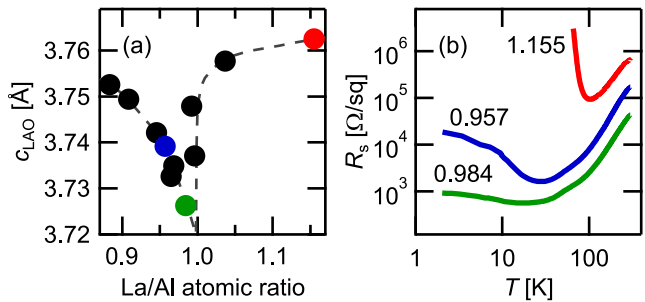}
\caption{(color online) (a) Out-of-plane lattice constant of the epitaxial LaAlO$_{3}$ films, as a function of the cation ratio obtained from the amorphous LaAlO$_{3}$ films grown at the same laser conditions. Dashed curve is a guide to the eye. (b) Temperature dependence of the sheet resistance of the LaAlO$_{3}$/SrTiO$_{3}$ samples with different film cation ratios. Numbers indicate the La/Al atomic ratio from ICP.}
\label{fig:fig1}
\end{figure}

LAO films were grown using a total of 11 different laser conditions, as summarized in Table~\ref{tb:tb1}, which also shows the cation ratios from ICP and $c_{\text{LAO}}$ from the XRD measurements.
The La/Al ratio was varied from 0.883 to 1.155, a range of more than $\pm$10{\%}, comparable to values reported for other similar systems.\cite{JApplPhys-86-6528,JApplPhys-103-103703,AdvMater-20-2528}
Two of the films were La rich, while the remaining were La poor to varying degrees, including two samples which were within 1{\%} of La/Al = 1.
As shown in Fig.~\ref{fig:fig1}(a), $c_{\text{LAO}}$ increases for both the La rich and La poor films, relative to La/Al = 1, where we estimate $c_{\text{LAO}} = $ 3.72 \text{ \AA}.

The estimated $c_{\text{LAO}}$ of the stoichiometric LAO film is in good agreement with calculations of the Poisson ratio,\cite{JApplPhys-104-073518} assuming a tensile strain by the STO substrate.\cite{RSM}
The $c_{\text{LAO}}$ expansion of the off-stoichiometric films is expected, since cation vacancies in insulating films cause lattice expansion due to Coulomb repulsion.\cite{JApplPhys-103-103703}
We note that the formation of interstitial or anti-sites defects is energetically unlikely.\cite{PhysRevB-80-104115}
Theoretical calculations show that the formation energy of ($V_{\text{Al}}^{\prime \prime \prime} + \frac{3}{2}V_{\text{O}}^{\bullet \bullet}$) is larger than that of ($V_{\text{La}}^{\prime \prime \prime} + \frac{3}{2}V_{\text{O}}^{\bullet \bullet}$) in a wide range of the La-O-Al ternary phase diagram\cite{PhysRevB-80-104115} (here we follow the notation of \citet{SolidStatePhys-3-307}), which qualitatively explains why most growth conditions result in La poor films.
It is also suggested that $V_{\text{Al}}^{\prime \prime \prime}$ produce larger atomic relaxations as compared to those from $V_{\text{La}}^{\prime \prime \prime}$,\cite{PhysRevB-80-104115} which may explain the asymmetry about La/Al = 1.
Interestingly, there are two films which are nearly stoichiometric from ICP, but show nontrivial expansion of $c_{\text{LAO}}$, which cannot be explained within the above scenario.
A likely interpretation is that both La and Al vacancies are kinetically induced in equal amounts at these conditions,\cite{PhysRevB-80-104115} and the oxygen content differs between the two.

We next examined the effect of the laser parameters on the electronic properties of the LAO/STO interface.
In Fig.~\ref{fig:fig1}(b) we show representative examples of the film stoichiometry dependence of the interface conductivity.
All of the samples show an upturn in the resistance at low temperatures ($T$), qualitatively consistent with previous reports.\cite{ApplPhysLett-94-222111,EurophysLett-91-17004}
With increasing the film off-stoichiometry, the sheet resistance becomes larger at all temperatures, and the upturn at low temperatures becomes more dramatic.
In the following we focus on the sheet carrier density ($n_{\text{s}}$) at 100 K, since at this temperature the resistance of all the samples was low enough to obtain clear Hall effect data sets, with an estimated maximum error bar of $\lesssim 10 \text{\%}$.
Also, the Hall effects at this temperature were fully linear to the highest measured fields (8 T).

\begin{figure}
\centering
\includegraphics{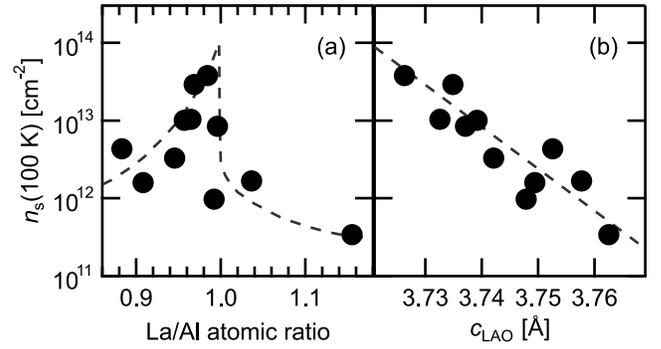}
\caption{Sheet carrier density of the LaAlO$_{3}$/SrTiO$_{3}$ samples at 100 K as a function of (a) the cation ratio from ICP and (b) the out-of-plane lattice constant of the LaAlO$_{3}$ films. Dashed curves are guides to the eye.}
\label{fig:fig2}
\end{figure}

Figures \ref{fig:fig2}(a) and (b) show $n_{\text{s}} (100 \text{ K})$ of the LAO/STO samples as a function of the LAO cation ratio from ICP and $c_{\text{LAO}}$, respectively.
There is a clear and significant relationship between the film stoichiometry, the lattice constant and the carrier density.
Despite some scatter, the samples with more off-stoichiometric films are found to have smaller $n_{\text{s}}$ [Fig.~\ref{fig:fig2}(a)].
Most notably, as shown in Fig.~\ref{fig:fig2}(b), all the carrier densities simply scaled with the film lattice constant over more than two orders of magnitude, independent of which cation(s) is (are) deficient.

\begin{figure*}
\centering
\includegraphics{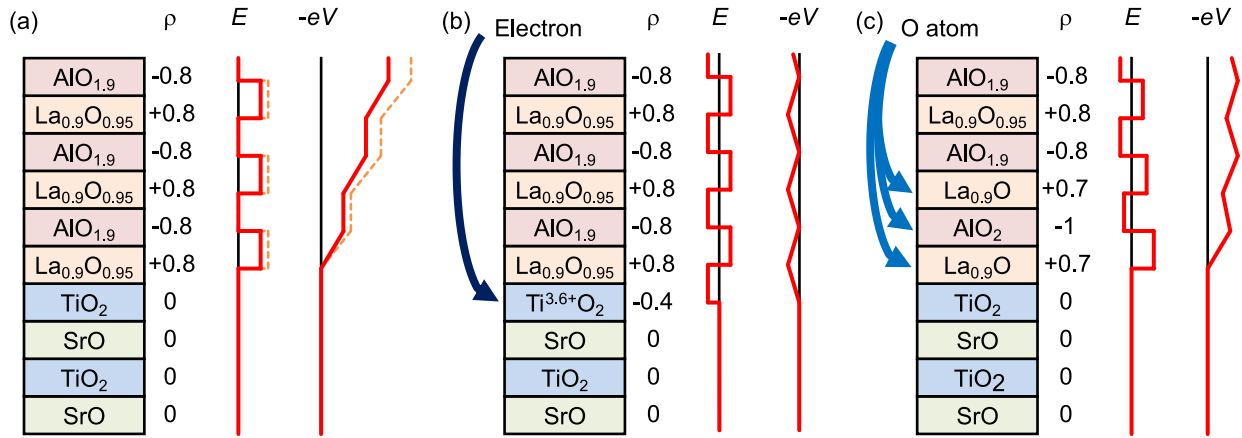}
\caption{(color online) Schematic structure, charge density $\rho$, electric field $E$ and electrostatic potential $-eV$ of the La$_{0.9}$AlO$_{3-\delta}$/SrTiO$_{3}$ $(001)$ heterostructure. (a) Unreconstructed structure with oxygen vacancies equally distributed amongst all oxygen sites. Dashed lines denote $E$ and $-eV$ of the stoichiometric structure. (b) Possible reconstruction via charge transfer, which induces less electrons than in the stoichometric structure. (c) Atomic reconstruction via stoichometry change, which resolves the polar problem without doping electrons to the substrate.}
\label{fig:fig3}
\end{figure*}

Next, we discuss the possible mechanisms that may explain these results.
The observed smaller carrier density for cation off-stoichiometric samples [Fig.~\ref{fig:fig2}] is not consistent with a simple picture of intermixing, since the film off-stoichiometry should enhance the cation diffusion from the substrate to the film, creating larger room for La ions to diffuse to the substrate, and enhance conductivity.
The data are also not consistent with a simple picture of oxygen vacancies by gettering from STO, since this would tend to increase with the density of vacancies caused by cation off-stoichiometry in LAO.
Similarly, blocking of reoxidation from molecular $\text{O}_{2}$ by the LAO should decrease as the LAO thickness decreases, while experimentally we find that at fixed laser conditions, thinner films show higher conductivity.
Thus we focus on the role of the electrostatic boundary conditions.

For simplicity, we consider the 10{\%} La deficient case, La$_{0.9}$AlO$_{3-\delta}$/SrTiO$_{3}$, as shown in Fig.~\ref{fig:fig3}.
Since La vacancies are negatively charged, it would be natural to assume that the off-stoichiometric LAO film also has positively charged oxygen vacancies in order to maintain bulk charge neutrality.
Despite the uncertainty in the exact density of the oxygen vacancies on each layer, here we illustrate the simplest case in which the oxygen vacancies are equally distributed amongst all the oxygen sites.
In this case, the absolute values of the charge density of each layer becomes smaller [Fig.~\ref{fig:fig3}(a)], similar to $(\text{LaAlO}_{3})_{x}(\text{SrTiO}_{3})_{1-x}$ solid solution films.\cite{arXiv:1112.3532}
Therefore, while there still remains a polar discontinuity, the extra charge required to resolve it is less than that in the stoichiometric structure, as schematically shown in Fig.~\ref{fig:fig3}(b).

Another important point is that the electronic reconstruction is not the only option for resolving the polar problem.
Atomic reconstruction is also possible, as in conventional semiconductor heterojunctions\cite{PhysRevB-18-4402} and at the $p$-type LAO/STO interface.\cite{NatureMater-5-204}
In the case of the $p$-type interface, the extra positive charge required for resolving the polar problem are provided by oxygen vacancies.\cite{NatureMater-5-204}
Analogous to this, the extra negative charge required at the $n$-type interface may be provided by \emph{excess} oxygen.
If the LAO film is fully stoichiometric, there is no room to accept the excess oxygen, given the tightly packed perovskite structure, and it is highly energetically unfavorable.\cite{PhysRevB-80-104115}
However, if the LAO film is significantly off-stoichiometric, it can accept the extra oxygen, as schematically shown in Fig.~\ref{fig:fig3}(c).
This atomic reconstruction resolves the polar problem without doping electrons to the STO substrate, resulting in an insulating interface.

These two schematic extremes of the possible reconstructions qualitatively explains why the film off-stoichiometry strongly reduces the sheet carrier density.
In particular, the atomic reconstruction model can explain the similar effect of La and Al vacancies [Fig.~\ref{fig:fig2}(a)]: since both cation vacancies require oxygen vacancies to maintain bulk charge neutrality, essentially the same atomic reconstruction is available in both cases.
However, it is difficult to quantitatively explain our observation within these extremal models.
In particular, purely within the electronic reconstruction model, a linear relationship between the cation ratio and $n_{\text{s}}$ is expected, which is not consistent with the data.
On the other hand, purely within the atomic reconstruction model, no conductivity is expected when the film is off-stoichiometric.
The experimentally realized situation is thus probably intermediate of these two extremes.
We hope that these results will motivate further theoretical study of defects and their energetics in off-stoichiometric heterostructures, in order to develop a clearer microscopic understanding of the results presented in this letter.

To summarize, we found a strong impact of the LAO film stoichiometry, modulated by the laser parameters, on the electronic properties of the LAO/STO interface.
In particular, film off-stoichiometry strongly reduced the carrier density at the interface.
We propose that film off-stoichiometry changes the balance between atomic and electronic reconstructions, the former resolving the polar discontinuity without doping electrons.
This film stoichiometry dependence provides new insights into the microscopic carrier generation mechanisms at the LAO/STO interface, and assists optimization of the electronic properties for future applications.
These results also gives better understanding of the growth dependence of the electronic phase diagrams of this system, which is vital for disentangling many of the conflicting studies by various groups. \\

We acknowledge Hitachi Kyowa Engineering Co., Ltd.\ for the ICP measurements.
This work was supported by the Department of Energy, Office of Basic Energy Sciences, Materials Sciences and Engineering Division, under contract DE-AC02-76SF00515.


\begin{thebibliography}{30}%
\makeatletter
\providecommand \@ifxundefined [1]{%
 \@ifx{#1\undefined}
}%
\providecommand \@ifnum [1]{%
 \ifnum #1\expandafter \@firstoftwo
 \else \expandafter \@secondoftwo
 \fi
}%
\providecommand \@ifx [1]{%
 \ifx #1\expandafter \@firstoftwo
 \else \expandafter \@secondoftwo
 \fi
}%
\providecommand \natexlab [1]{#1}%
\providecommand \enquote  [1]{``#1''}%
\providecommand \bibnamefont  [1]{#1}%
\providecommand \bibfnamefont [1]{#1}%
\providecommand \citenamefont [1]{#1}%
\providecommand \href@noop [0]{\@secondoftwo}%
\providecommand \href [0]{\begingroup \@sanitize@url \@href}%
\providecommand \@href[1]{\@@startlink{#1}\@@href}%
\providecommand \@@href[1]{\endgroup#1\@@endlink}%
\providecommand \@sanitize@url [0]{\catcode `\\12\catcode `\$12\catcode
  `\&12\catcode `\#12\catcode `\^12\catcode `\_12\catcode `\%12\relax}%
\providecommand \@@startlink[1]{}%
\providecommand \@@endlink[0]{}%
\providecommand \url  [0]{\begingroup\@sanitize@url \@url }%
\providecommand \@url [1]{\endgroup\@href {#1}{\urlprefix }}%
\providecommand \urlprefix  [0]{URL }%
\providecommand \Eprint [0]{\href }%
\providecommand \doibase [0]{http://dx.doi.org/}%
\providecommand \selectlanguage [0]{\@gobble}%
\providecommand \bibinfo  [0]{\@secondoftwo}%
\providecommand \bibfield  [0]{\@secondoftwo}%
\providecommand \translation [1]{[#1]}%
\providecommand \BibitemOpen [0]{}%
\providecommand \bibitemStop [0]{}%
\providecommand \bibitemNoStop [0]{.\EOS\space}%
\providecommand \EOS [0]{\spacefactor3000\relax}%
\providecommand \BibitemShut  [1]{\csname bibitem#1\endcsname}%
\let\auto@bib@innerbib\@empty
\bibitem [{\citenamefont {Ohtomo}\ and\ \citenamefont
  {Hwang}(2004)}]{Nature-427-423}%
  \BibitemOpen
  \bibfield  {author} {\bibinfo {author} {\bibfnamefont {A.}~\bibnamefont
  {Ohtomo}}\ and\ \bibinfo {author} {\bibfnamefont {H.~Y.}\ \bibnamefont
  {Hwang}},\ }\href {\doibase 10.1038/nature02308} {\bibfield  {journal}
  {\bibinfo  {journal} {Nature (London)}\ }\textbf {\bibinfo {volume} {427}},\
  \bibinfo {pages} {423} (\bibinfo {year} {2004})}\BibitemShut {NoStop}%
\bibitem [{\citenamefont {Nakagawa}, \citenamefont {Hwang},\ and\ \citenamefont
  {Muller}(2006)}]{NatureMater-5-204}%
  \BibitemOpen
  \bibfield  {author} {\bibinfo {author} {\bibfnamefont {N.}~\bibnamefont
  {Nakagawa}}, \bibinfo {author} {\bibfnamefont {H.~Y.}\ \bibnamefont {Hwang}},
  \ and\ \bibinfo {author} {\bibfnamefont {D.~A.}\ \bibnamefont {Muller}},\
  }\href {\doibase 10.1038/nmat1569} {\bibfield  {journal} {\bibinfo  {journal}
  {Nature Mater.}\ }\textbf {\bibinfo {volume} {5}},\ \bibinfo {pages} {204}
  (\bibinfo {year} {2006})}\BibitemShut {NoStop}%
\bibitem [{\citenamefont {Kalabukhov}\ \emph {et~al.}(2007)\citenamefont
  {Kalabukhov}, \citenamefont {Gunnarsson}, \citenamefont {B{\"{o}}rjesson},
  \citenamefont {Olsson}, \citenamefont {Claeson},\ and\ \citenamefont
  {Winkler}}]{PhysRevB-75-121404}%
  \BibitemOpen
  \bibfield  {author} {\bibinfo {author} {\bibfnamefont {A.}~\bibnamefont
  {Kalabukhov}}, \bibinfo {author} {\bibfnamefont {R.}~\bibnamefont
  {Gunnarsson}}, \bibinfo {author} {\bibfnamefont {J.}~\bibnamefont
  {B{\"{o}}rjesson}}, \bibinfo {author} {\bibfnamefont {E.}~\bibnamefont
  {Olsson}}, \bibinfo {author} {\bibfnamefont {T.}~\bibnamefont {Claeson}}, \
  and\ \bibinfo {author} {\bibfnamefont {D.}~\bibnamefont {Winkler}},\ }\href
  {\doibase 10.1103/PhysRevB.75.121404} {\bibfield  {journal} {\bibinfo
  {journal} {Phys. Rev. B}\ }\textbf {\bibinfo {volume} {75}},\ \bibinfo {eid}
  {121404} (\bibinfo {year} {2007})}\BibitemShut {NoStop}%
\bibitem [{\citenamefont {Siemons}\ \emph {et~al.}(2007)\citenamefont
  {Siemons}, \citenamefont {Koster}, \citenamefont {Yamamoto}, \citenamefont
  {Harrison}, \citenamefont {Lucovsky}, \citenamefont {Geballe}, \citenamefont
  {Blank},\ and\ \citenamefont {Beasley}}]{PhysRevLett-98-196802}%
  \BibitemOpen
  \bibfield  {author} {\bibinfo {author} {\bibfnamefont {W.}~\bibnamefont
  {Siemons}}, \bibinfo {author} {\bibfnamefont {G.}~\bibnamefont {Koster}},
  \bibinfo {author} {\bibfnamefont {H.}~\bibnamefont {Yamamoto}}, \bibinfo
  {author} {\bibfnamefont {W.~A.}\ \bibnamefont {Harrison}}, \bibinfo {author}
  {\bibfnamefont {G.}~\bibnamefont {Lucovsky}}, \bibinfo {author}
  {\bibfnamefont {T.~H.}\ \bibnamefont {Geballe}}, \bibinfo {author}
  {\bibfnamefont {D.~H.~A.}\ \bibnamefont {Blank}}, \ and\ \bibinfo {author}
  {\bibfnamefont {M.~R.}\ \bibnamefont {Beasley}},\ }\href {\doibase
  10.1103/PhysRevLett.98.196802} {\bibfield  {journal} {\bibinfo  {journal}
  {Phys. Rev. Lett.}\ }\textbf {\bibinfo {volume} {98}},\ \bibinfo {eid}
  {196802} (\bibinfo {year} {2007})}\BibitemShut {NoStop}%
\bibitem [{\citenamefont {Herranz}\ \emph {et~al.}(2007)\citenamefont
  {Herranz}, \citenamefont {Basleti{\'{c}}}, \citenamefont {Bibes},
  \citenamefont {Carr{\'{e}}t{\'{e}}ro}, \citenamefont {Tafra}, \citenamefont
  {Jacquet}, \citenamefont {Bouzehouane}, \citenamefont {Deranlot},
  \citenamefont {Hamzi{\'{c}}}, \citenamefont {Broto}, \citenamefont
  {Barth{\'{e}}l{\'{e}}my},\ and\ \citenamefont
  {Fert}}]{PhysRevLett-98-216803}%
  \BibitemOpen
  \bibfield  {author} {\bibinfo {author} {\bibfnamefont {G.}~\bibnamefont
  {Herranz}}, \bibinfo {author} {\bibfnamefont {M.}~\bibnamefont
  {Basleti{\'{c}}}}, \bibinfo {author} {\bibfnamefont {M.}~\bibnamefont
  {Bibes}}, \bibinfo {author} {\bibfnamefont {C.}~\bibnamefont
  {Carr{\'{e}}t{\'{e}}ro}}, \bibinfo {author} {\bibfnamefont {E.}~\bibnamefont
  {Tafra}}, \bibinfo {author} {\bibfnamefont {E.}~\bibnamefont {Jacquet}},
  \bibinfo {author} {\bibfnamefont {K.}~\bibnamefont {Bouzehouane}}, \bibinfo
  {author} {\bibfnamefont {C.}~\bibnamefont {Deranlot}}, \bibinfo {author}
  {\bibfnamefont {A.}~\bibnamefont {Hamzi{\'{c}}}}, \bibinfo {author}
  {\bibfnamefont {J.-M.}\ \bibnamefont {Broto}}, \bibinfo {author}
  {\bibfnamefont {A.}~\bibnamefont {Barth{\'{e}}l{\'{e}}my}}, \ and\ \bibinfo
  {author} {\bibfnamefont {A.}~\bibnamefont {Fert}},\ }\href {\doibase
  10.1103/PhysRevLett.98.216803} {\bibfield  {journal} {\bibinfo  {journal}
  {Phys. Rev. Lett.}\ }\textbf {\bibinfo {volume} {98}},\ \bibinfo {eid}
  {216803} (\bibinfo {year} {2007})}\BibitemShut {NoStop}%
\bibitem [{\citenamefont {Willmott}\ \emph {et~al.}(2007)\citenamefont
  {Willmott}, \citenamefont {Pauli}, \citenamefont {Herger}, \citenamefont
  {Schlep{\"{u}}tz}, \citenamefont {Martoccia}, \citenamefont {Patterson},
  \citenamefont {Delley}, \citenamefont {Clarke}, \citenamefont {Kumah},
  \citenamefont {Cionca},\ and\ \citenamefont
  {Yacoby}}]{PhysRevLett-99-155502}%
  \BibitemOpen
  \bibfield  {author} {\bibinfo {author} {\bibfnamefont {P.~R.}\ \bibnamefont
  {Willmott}}, \bibinfo {author} {\bibfnamefont {S.~A.}\ \bibnamefont {Pauli}},
  \bibinfo {author} {\bibfnamefont {R.}~\bibnamefont {Herger}}, \bibinfo
  {author} {\bibfnamefont {C.~M.}\ \bibnamefont {Schlep{\"{u}}tz}}, \bibinfo
  {author} {\bibfnamefont {D.}~\bibnamefont {Martoccia}}, \bibinfo {author}
  {\bibfnamefont {B.~D.}\ \bibnamefont {Patterson}}, \bibinfo {author}
  {\bibfnamefont {B.}~\bibnamefont {Delley}}, \bibinfo {author} {\bibfnamefont
  {R.}~\bibnamefont {Clarke}}, \bibinfo {author} {\bibfnamefont
  {D.}~\bibnamefont {Kumah}}, \bibinfo {author} {\bibfnamefont
  {C.}~\bibnamefont {Cionca}}, \ and\ \bibinfo {author} {\bibfnamefont
  {Y.}~\bibnamefont {Yacoby}},\ }\href {\doibase 10.1103/PhysRevLett.99.155502}
  {\bibfield  {journal} {\bibinfo  {journal} {Phys. Rev. Lett.}\ }\textbf
  {\bibinfo {volume} {99}},\ \bibinfo {eid} {155502} (\bibinfo {year}
  {2007})}\BibitemShut {NoStop}%
\bibitem [{\citenamefont {Chambers}\ \emph {et~al.}(2010)\citenamefont
  {Chambers}, \citenamefont {Engelhard}, \citenamefont {Shutthanandan},
  \citenamefont {Zhu}, \citenamefont {Droubay}, \citenamefont {Qiao},
  \citenamefont {Sushko}, \citenamefont {Feng}, \citenamefont {Lee},
  \citenamefont {Gustafsson}, \citenamefont {Garfunkel}, \citenamefont {Shah},
  \citenamefont {Zuo},\ and\ \citenamefont {Ramasse}}]{SurfSciRep-65-317}%
  \BibitemOpen
  \bibfield  {author} {\bibinfo {author} {\bibfnamefont {S.~A.}\ \bibnamefont
  {Chambers}}, \bibinfo {author} {\bibfnamefont {M.~H.}\ \bibnamefont
  {Engelhard}}, \bibinfo {author} {\bibfnamefont {V.}~\bibnamefont
  {Shutthanandan}}, \bibinfo {author} {\bibfnamefont {Z.}~\bibnamefont {Zhu}},
  \bibinfo {author} {\bibfnamefont {T.~C.}\ \bibnamefont {Droubay}}, \bibinfo
  {author} {\bibfnamefont {L.}~\bibnamefont {Qiao}}, \bibinfo {author}
  {\bibfnamefont {P.~V.}\ \bibnamefont {Sushko}}, \bibinfo {author}
  {\bibfnamefont {T.}~\bibnamefont {Feng}}, \bibinfo {author} {\bibfnamefont
  {H.~D.}\ \bibnamefont {Lee}}, \bibinfo {author} {\bibfnamefont
  {T.}~\bibnamefont {Gustafsson}}, \bibinfo {author} {\bibfnamefont
  {E.}~\bibnamefont {Garfunkel}}, \bibinfo {author} {\bibfnamefont {A.~B.}\
  \bibnamefont {Shah}}, \bibinfo {author} {\bibfnamefont {J.-M.}\ \bibnamefont
  {Zuo}}, \ and\ \bibinfo {author} {\bibfnamefont {Q.~M.}\ \bibnamefont
  {Ramasse}},\ }\href {\doibase DOI: 10.1016/j.surfrep.2010.09.001} {\bibfield
  {journal} {\bibinfo  {journal} {Surf. Sci. Rep.}\ }\textbf {\bibinfo {volume}
  {65}},\ \bibinfo {pages} {317} (\bibinfo {year} {2010})}\BibitemShut
  {NoStop}%
\bibitem [{\citenamefont {Higuchi}\ and\ \citenamefont
  {Hwang}()}]{arXiv:1105.5779}%
  \BibitemOpen
  \bibfield  {author} {\bibinfo {author} {\bibfnamefont {T.}~\bibnamefont
  {Higuchi}}\ and\ \bibinfo {author} {\bibfnamefont {H.~Y.}\ \bibnamefont
  {Hwang}},\ }\href@noop {} {}\bibinfo {note} {``General Considerations of the
  Electrostatic Boundary Conditions in Oxide Heterostructures,'' in
  \emph{Multifunctional Oxide Heterostructures}, edited by E. Tsymbal, C. B.
  Eom, R. Ramesh, and E. Dagotto, pp. 183-213, Oxford University Press
  (2012)}\BibitemShut {NoStop}%
\bibitem [{\citenamefont {Brinkman}\ \emph {et~al.}(2007)\citenamefont
  {Brinkman}, \citenamefont {Huijben}, \citenamefont {van Zalk}, \citenamefont
  {Huijben}, \citenamefont {Zeitler}, \citenamefont {Maan}, \citenamefont
  {van~der Wiel}, \citenamefont {Rijnders}, \citenamefont {Blank},\ and\
  \citenamefont {Hilgenkamp}}]{NatureMater-6-493}%
  \BibitemOpen
  \bibfield  {author} {\bibinfo {author} {\bibfnamefont {A.}~\bibnamefont
  {Brinkman}}, \bibinfo {author} {\bibfnamefont {M.}~\bibnamefont {Huijben}},
  \bibinfo {author} {\bibfnamefont {M.}~\bibnamefont {van Zalk}}, \bibinfo
  {author} {\bibfnamefont {J.}~\bibnamefont {Huijben}}, \bibinfo {author}
  {\bibfnamefont {U.}~\bibnamefont {Zeitler}}, \bibinfo {author} {\bibfnamefont
  {J.~C.}\ \bibnamefont {Maan}}, \bibinfo {author} {\bibfnamefont {W.~G.}\
  \bibnamefont {van~der Wiel}}, \bibinfo {author} {\bibfnamefont
  {G.}~\bibnamefont {Rijnders}}, \bibinfo {author} {\bibfnamefont {D.~H.~A.}\
  \bibnamefont {Blank}}, \ and\ \bibinfo {author} {\bibfnamefont
  {H.}~\bibnamefont {Hilgenkamp}},\ }\href {\doibase 10.1038/nmat1931}
  {\bibfield  {journal} {\bibinfo  {journal} {Nature Mater.}\ }\textbf
  {\bibinfo {volume} {6}},\ \bibinfo {pages} {493} (\bibinfo {year}
  {2007})}\BibitemShut {NoStop}%
\bibitem [{\citenamefont {Basletic}\ \emph {et~al.}(2008)\citenamefont
  {Basletic}, \citenamefont {Maurice}, \citenamefont {Carr{\'{e}}t{\'{e}}ro},
  \citenamefont {Herranz}, \citenamefont {Copie}, \citenamefont {Bibes},
  \citenamefont {Jacquet}, \citenamefont {Bouzehouane}, \citenamefont {Fusil},\
  and\ \citenamefont {Barth{\'{e}}l{\'{e}}my}}]{NatureMater-7-621}%
  \BibitemOpen
  \bibfield  {author} {\bibinfo {author} {\bibfnamefont {M.}~\bibnamefont
  {Basletic}}, \bibinfo {author} {\bibfnamefont {J.-L.}\ \bibnamefont
  {Maurice}}, \bibinfo {author} {\bibfnamefont {C.}~\bibnamefont
  {Carr{\'{e}}t{\'{e}}ro}}, \bibinfo {author} {\bibfnamefont {G.}~\bibnamefont
  {Herranz}}, \bibinfo {author} {\bibfnamefont {O.}~\bibnamefont {Copie}},
  \bibinfo {author} {\bibfnamefont {M.}~\bibnamefont {Bibes}}, \bibinfo
  {author} {\bibfnamefont {{\'{E}}.}~\bibnamefont {Jacquet}}, \bibinfo {author}
  {\bibfnamefont {K.}~\bibnamefont {Bouzehouane}}, \bibinfo {author}
  {\bibfnamefont {S.}~\bibnamefont {Fusil}}, \ and\ \bibinfo {author}
  {\bibfnamefont {A.}~\bibnamefont {Barth{\'{e}}l{\'{e}}my}},\ }\href {\doibase
  10.1038/nmat2223} {\bibfield  {journal} {\bibinfo  {journal} {Nature Mater.}\
  }\textbf {\bibinfo {volume} {7}},\ \bibinfo {pages} {621} (\bibinfo {year}
  {2008})}\BibitemShut {NoStop}%
\bibitem [{\citenamefont {Cancellieri}\ \emph {et~al.}(2010)\citenamefont
  {Cancellieri}, \citenamefont {Reyren}, \citenamefont {Gariglio},
  \citenamefont {Caviglia}, \citenamefont {Fete},\ and\ \citenamefont
  {Triscone}}]{EurophysLett-91-17004}%
  \BibitemOpen
  \bibfield  {author} {\bibinfo {author} {\bibfnamefont {C.}~\bibnamefont
  {Cancellieri}}, \bibinfo {author} {\bibfnamefont {N.}~\bibnamefont {Reyren}},
  \bibinfo {author} {\bibfnamefont {S.}~\bibnamefont {Gariglio}}, \bibinfo
  {author} {\bibfnamefont {A.~D.}\ \bibnamefont {Caviglia}}, \bibinfo {author}
  {\bibfnamefont {A.}~\bibnamefont {Fete}}, \ and\ \bibinfo {author}
  {\bibfnamefont {J.-M.}\ \bibnamefont {Triscone}},\ }\href@noop {} {\bibfield
  {journal} {\bibinfo  {journal} {EPL}\ }\textbf {\bibinfo {volume} {91}},\
  \bibinfo {pages} {17004} (\bibinfo {year} {2010})}\BibitemShut {NoStop}%
\bibitem [{\citenamefont {Caviglia}\ \emph {et~al.}(2010)\citenamefont
  {Caviglia}, \citenamefont {Gariglio}, \citenamefont {Cancellieri},
  \citenamefont {Sac{\'{e}}p{\'{e}}}, \citenamefont {F{\^{e}}te}, \citenamefont
  {Reyren}, \citenamefont {Gabay}, \citenamefont {Morpurgo},\ and\
  \citenamefont {Triscone}}]{PhysRevLett-105-236802}%
  \BibitemOpen
  \bibfield  {author} {\bibinfo {author} {\bibfnamefont {A.~D.}\ \bibnamefont
  {Caviglia}}, \bibinfo {author} {\bibfnamefont {S.}~\bibnamefont {Gariglio}},
  \bibinfo {author} {\bibfnamefont {C.}~\bibnamefont {Cancellieri}}, \bibinfo
  {author} {\bibfnamefont {B.}~\bibnamefont {Sac{\'{e}}p{\'{e}}}}, \bibinfo
  {author} {\bibfnamefont {A.}~\bibnamefont {F{\^{e}}te}}, \bibinfo {author}
  {\bibfnamefont {N.}~\bibnamefont {Reyren}}, \bibinfo {author} {\bibfnamefont
  {M.}~\bibnamefont {Gabay}}, \bibinfo {author} {\bibfnamefont {A.~F.}\
  \bibnamefont {Morpurgo}}, \ and\ \bibinfo {author} {\bibfnamefont {J.-M.}\
  \bibnamefont {Triscone}},\ }\href {\doibase 10.1103/PhysRevLett.105.236802}
  {\bibfield  {journal} {\bibinfo  {journal} {Phys. Rev. Lett.}\ }\textbf
  {\bibinfo {volume} {105}},\ \bibinfo {pages} {236802} (\bibinfo {year}
  {2010})}\BibitemShut {NoStop}%
\bibitem [{\citenamefont {Thiel}\ \emph {et~al.}(2006)\citenamefont {Thiel},
  \citenamefont {Hammerl}, \citenamefont {Schmehl}, \citenamefont {Schneider},\
  and\ \citenamefont {Mannhart}}]{Science-313-1942}%
  \BibitemOpen
  \bibfield  {author} {\bibinfo {author} {\bibfnamefont {S.}~\bibnamefont
  {Thiel}}, \bibinfo {author} {\bibfnamefont {G.}~\bibnamefont {Hammerl}},
  \bibinfo {author} {\bibfnamefont {A.}~\bibnamefont {Schmehl}}, \bibinfo
  {author} {\bibfnamefont {C.~W.}\ \bibnamefont {Schneider}}, \ and\ \bibinfo
  {author} {\bibfnamefont {J.}~\bibnamefont {Mannhart}},\ }\href {\doibase
  10.1126/science.1131091} {\bibfield  {journal} {\bibinfo  {journal}
  {Science}\ }\textbf {\bibinfo {volume} {313}},\ \bibinfo {pages} {1942}
  (\bibinfo {year} {2006})}\BibitemShut {NoStop}%
\bibitem [{\citenamefont {Bell}\ \emph {et~al.}(2009)\citenamefont {Bell},
  \citenamefont {Harashima}, \citenamefont {Hikita},\ and\ \citenamefont
  {Hwang}}]{ApplPhysLett-94-222111}%
  \BibitemOpen
  \bibfield  {author} {\bibinfo {author} {\bibfnamefont {C.}~\bibnamefont
  {Bell}}, \bibinfo {author} {\bibfnamefont {S.}~\bibnamefont {Harashima}},
  \bibinfo {author} {\bibfnamefont {Y.}~\bibnamefont {Hikita}}, \ and\ \bibinfo
  {author} {\bibfnamefont {H.~Y.}\ \bibnamefont {Hwang}},\ }\href {\doibase
  10.1063/1.3149695} {\bibfield  {journal} {\bibinfo  {journal} {Appl. Phys.
  Lett.}\ }\textbf {\bibinfo {volume} {94}},\ \bibinfo {eid} {222111} (\bibinfo
  {year} {2009})}\BibitemShut {NoStop}%
\bibitem [{\citenamefont {Nishimura}\ \emph {et~al.}(2004)\citenamefont
  {Nishimura}, \citenamefont {Ohtomo}, \citenamefont {Ohkubo}, \citenamefont
  {Murakami},\ and\ \citenamefont {Kawasaki}}]{JpnJApplPhys-43-L1032}%
  \BibitemOpen
  \bibfield  {author} {\bibinfo {author} {\bibfnamefont {J.}~\bibnamefont
  {Nishimura}}, \bibinfo {author} {\bibfnamefont {A.}~\bibnamefont {Ohtomo}},
  \bibinfo {author} {\bibfnamefont {A.}~\bibnamefont {Ohkubo}}, \bibinfo
  {author} {\bibfnamefont {Y.}~\bibnamefont {Murakami}}, \ and\ \bibinfo
  {author} {\bibfnamefont {M.}~\bibnamefont {Kawasaki}},\ }\href {\doibase
  10.1143/JJAP.43.L1032} {\bibfield  {journal} {\bibinfo  {journal} {Jpn. J.
  Appl. Phys.}\ }\textbf {\bibinfo {volume} {43}},\ \bibinfo {pages} {L1032}
  (\bibinfo {year} {2004})}\BibitemShut {NoStop}%
\bibitem [{\citenamefont {Yamamoto}\ \emph {et~al.}(2011)\citenamefont
  {Yamamoto}, \citenamefont {Bell}, \citenamefont {Hikita}, \citenamefont
  {Hwang}, \citenamefont {Nakamura}, \citenamefont {Kimura},\ and\
  \citenamefont {Wakabayashi}}]{PhysRevLett-107-036104}%
  \BibitemOpen
  \bibfield  {author} {\bibinfo {author} {\bibfnamefont {R.}~\bibnamefont
  {Yamamoto}}, \bibinfo {author} {\bibfnamefont {C.}~\bibnamefont {Bell}},
  \bibinfo {author} {\bibfnamefont {Y.}~\bibnamefont {Hikita}}, \bibinfo
  {author} {\bibfnamefont {H.~Y.}\ \bibnamefont {Hwang}}, \bibinfo {author}
  {\bibfnamefont {H.}~\bibnamefont {Nakamura}}, \bibinfo {author}
  {\bibfnamefont {T.}~\bibnamefont {Kimura}}, \ and\ \bibinfo {author}
  {\bibfnamefont {Y.}~\bibnamefont {Wakabayashi}},\ }\href {\doibase
  10.1103/PhysRevLett.107.036104} {\bibfield  {journal} {\bibinfo  {journal}
  {Phys. Rev. Lett.}\ }\textbf {\bibinfo {volume} {107}},\ \bibinfo {pages}
  {036104} (\bibinfo {year} {2011})}\BibitemShut {NoStop}%
\bibitem [{\citenamefont {Huijbregtse}\ \emph {et~al.}(1999)\citenamefont
  {Huijbregtse}, \citenamefont {Dam}, \citenamefont {Rector},\ and\
  \citenamefont {Griessen}}]{JApplPhys-86-6528}%
  \BibitemOpen
  \bibfield  {author} {\bibinfo {author} {\bibfnamefont {J.~M.}\ \bibnamefont
  {Huijbregtse}}, \bibinfo {author} {\bibfnamefont {B.}~\bibnamefont {Dam}},
  \bibinfo {author} {\bibfnamefont {J.~H.}\ \bibnamefont {Rector}}, \ and\
  \bibinfo {author} {\bibfnamefont {R.}~\bibnamefont {Griessen}},\ }\href
  {\doibase 10.1063/1.371619} {\bibfield  {journal} {\bibinfo  {journal} {J.
  Appl. Phys.}\ }\textbf {\bibinfo {volume} {86}},\ \bibinfo {pages} {6528}
  (\bibinfo {year} {1999})}\BibitemShut {NoStop}%
\bibitem [{\citenamefont {Ohnishi}\ \emph {et~al.}(2008)\citenamefont
  {Ohnishi}, \citenamefont {Shibuya}, \citenamefont {Yamamoto},\ and\
  \citenamefont {Lippmaa}}]{JApplPhys-103-103703}%
  \BibitemOpen
  \bibfield  {author} {\bibinfo {author} {\bibfnamefont {T.}~\bibnamefont
  {Ohnishi}}, \bibinfo {author} {\bibfnamefont {K.}~\bibnamefont {Shibuya}},
  \bibinfo {author} {\bibfnamefont {T.}~\bibnamefont {Yamamoto}}, \ and\
  \bibinfo {author} {\bibfnamefont {M.}~\bibnamefont {Lippmaa}},\ }\href
  {\doibase 10.1063/1.2921972} {\bibfield  {journal} {\bibinfo  {journal} {J.
  Appl. Phys.}\ }\textbf {\bibinfo {volume} {103}},\ \bibinfo {eid} {103703}
  (\bibinfo {year} {2008})}\BibitemShut {NoStop}%
\bibitem [{\citenamefont {Song}, \citenamefont {Susaki},\ and\ \citenamefont
  {Hwang}(2008)}]{AdvMater-20-2528}%
  \BibitemOpen
  \bibfield  {author} {\bibinfo {author} {\bibfnamefont {J.~H.}\ \bibnamefont
  {Song}}, \bibinfo {author} {\bibfnamefont {T.}~\bibnamefont {Susaki}}, \ and\
  \bibinfo {author} {\bibfnamefont {H.~Y.}\ \bibnamefont {Hwang}},\ }\href
  {\doibase 10.1002/adma.200701919} {\bibfield  {journal} {\bibinfo  {journal}
  {Adv. Mater.}\ }\textbf {\bibinfo {volume} {20}},\ \bibinfo {pages} {2528}
  (\bibinfo {year} {2008})}\BibitemShut {NoStop}%
\bibitem [{\citenamefont {Qiao}\ \emph {et~al.}(2011)\citenamefont {Qiao},
  \citenamefont {Droubay}, \citenamefont {Varga}, \citenamefont {Bowden},
  \citenamefont {Shutthanandan}, \citenamefont {Zhu}, \citenamefont {Kaspar},\
  and\ \citenamefont {Chambers}}]{PhysRevB-83-085408}%
  \BibitemOpen
  \bibfield  {author} {\bibinfo {author} {\bibfnamefont {L.}~\bibnamefont
  {Qiao}}, \bibinfo {author} {\bibfnamefont {T.~C.}\ \bibnamefont {Droubay}},
  \bibinfo {author} {\bibfnamefont {T.}~\bibnamefont {Varga}}, \bibinfo
  {author} {\bibfnamefont {M.~E.}\ \bibnamefont {Bowden}}, \bibinfo {author}
  {\bibfnamefont {V.}~\bibnamefont {Shutthanandan}}, \bibinfo {author}
  {\bibfnamefont {Z.}~\bibnamefont {Zhu}}, \bibinfo {author} {\bibfnamefont
  {T.~C.}\ \bibnamefont {Kaspar}}, \ and\ \bibinfo {author} {\bibfnamefont
  {S.~A.}\ \bibnamefont {Chambers}},\ }\href {\doibase
  10.1103/PhysRevB.83.085408} {\bibfield  {journal} {\bibinfo  {journal} {Phys.
  Rev. B}\ }\textbf {\bibinfo {volume} {83}},\ \bibinfo {pages} {085408}
  (\bibinfo {year} {2011})}\BibitemShut {NoStop}%
\bibitem [{\citenamefont {Schoofs}\ \emph {et~al.}(2011)\citenamefont
  {Schoofs}, \citenamefont {Fix}, \citenamefont {Kalabukhov}, \citenamefont
  {Winkler}, \citenamefont {Boikov}, \citenamefont {Serenkov}, \citenamefont
  {Sakharov}, \citenamefont {Claeson}, \citenamefont {MacManus-Driscoll},\ and\
  \citenamefont {Blamire}}]{JPhysCondensMatter-23-305002}%
  \BibitemOpen
  \bibfield  {author} {\bibinfo {author} {\bibfnamefont {F.}~\bibnamefont
  {Schoofs}}, \bibinfo {author} {\bibfnamefont {T.}~\bibnamefont {Fix}},
  \bibinfo {author} {\bibfnamefont {A.~S.}\ \bibnamefont {Kalabukhov}},
  \bibinfo {author} {\bibfnamefont {D.}~\bibnamefont {Winkler}}, \bibinfo
  {author} {\bibfnamefont {Y.}~\bibnamefont {Boikov}}, \bibinfo {author}
  {\bibfnamefont {I.}~\bibnamefont {Serenkov}}, \bibinfo {author}
  {\bibfnamefont {V.}~\bibnamefont {Sakharov}}, \bibinfo {author}
  {\bibfnamefont {T.}~\bibnamefont {Claeson}}, \bibinfo {author} {\bibfnamefont
  {J.~L.}\ \bibnamefont {MacManus-Driscoll}}, \ and\ \bibinfo {author}
  {\bibfnamefont {M.~G.}\ \bibnamefont {Blamire}},\ }\href@noop {} {\bibfield
  {journal} {\bibinfo  {journal} {J. Phys.: Condens. Matter}\ }\textbf
  {\bibinfo {volume} {23}},\ \bibinfo {pages} {305002} (\bibinfo {year}
  {2011})}\BibitemShut {NoStop}%
\bibitem [{\citenamefont {Kawasaki}\ \emph {et~al.}(1994)\citenamefont
  {Kawasaki}, \citenamefont {Takahashi}, \citenamefont {Maeda}, \citenamefont
  {Tsuchiya}, \citenamefont {Shinohara}, \citenamefont {Ishiyama},
  \citenamefont {Yonezawa}, \citenamefont {Yoshimoto},\ and\ \citenamefont
  {Koinuma}}]{Science-266-1540}%
  \BibitemOpen
  \bibfield  {author} {\bibinfo {author} {\bibfnamefont {M.}~\bibnamefont
  {Kawasaki}}, \bibinfo {author} {\bibfnamefont {K.}~\bibnamefont {Takahashi}},
  \bibinfo {author} {\bibfnamefont {T.}~\bibnamefont {Maeda}}, \bibinfo
  {author} {\bibfnamefont {R.}~\bibnamefont {Tsuchiya}}, \bibinfo {author}
  {\bibfnamefont {M.}~\bibnamefont {Shinohara}}, \bibinfo {author}
  {\bibfnamefont {O.}~\bibnamefont {Ishiyama}}, \bibinfo {author}
  {\bibfnamefont {T.}~\bibnamefont {Yonezawa}}, \bibinfo {author}
  {\bibfnamefont {M.}~\bibnamefont {Yoshimoto}}, \ and\ \bibinfo {author}
  {\bibfnamefont {H.}~\bibnamefont {Koinuma}},\ }\href {\doibase
  10.1126/science.266.5190.1540} {\bibfield  {journal} {\bibinfo  {journal}
  {Science}\ }\textbf {\bibinfo {volume} {266}},\ \bibinfo {pages} {1540}
  (\bibinfo {year} {1994})}\BibitemShut {NoStop}%
\bibitem [{\citenamefont {Koster}\ \emph {et~al.}(1998)\citenamefont {Koster},
  \citenamefont {Kropman}, \citenamefont {Rijnders}, \citenamefont {Blank},\
  and\ \citenamefont {Rogalla}}]{ApplPhysLett-73-2920}%
  \BibitemOpen
  \bibfield  {author} {\bibinfo {author} {\bibfnamefont {G.}~\bibnamefont
  {Koster}}, \bibinfo {author} {\bibfnamefont {B.~L.}\ \bibnamefont {Kropman}},
  \bibinfo {author} {\bibfnamefont {G.~J. H.~M.}\ \bibnamefont {Rijnders}},
  \bibinfo {author} {\bibfnamefont {D.~H.~A.}\ \bibnamefont {Blank}}, \ and\
  \bibinfo {author} {\bibfnamefont {H.}~\bibnamefont {Rogalla}},\ }\href
  {\doibase 10.1063/1.122630} {\bibfield  {journal} {\bibinfo  {journal} {Appl.
  Phys. Lett.}\ }\textbf {\bibinfo {volume} {73}},\ \bibinfo {pages} {2920}
  (\bibinfo {year} {1998})}\BibitemShut {NoStop}%
\bibitem [{LAO()}]{LAOindex}%
  \BibitemOpen
  \href@noop {} {}\bibinfo {note} {In this letter, all the {M}iller indices and
  the lattice constants of {LAO} are given for the pseudo-cubic perovskite
  unit.}\BibitemShut {Stop}%
\bibitem [{\citenamefont {Luo}\ and\ \citenamefont
  {Wang}(2008)}]{JApplPhys-104-073518}%
  \BibitemOpen
  \bibfield  {author} {\bibinfo {author} {\bibfnamefont {X.}~\bibnamefont
  {Luo}}\ and\ \bibinfo {author} {\bibfnamefont {B.}~\bibnamefont {Wang}},\
  }\href {\doibase 10.1063/1.2990068} {\bibfield  {journal} {\bibinfo
  {journal} {J. Appl. Phys.}\ }\textbf {\bibinfo {volume} {104}},\ \bibinfo
  {eid} {073518} (\bibinfo {year} {2008})}\BibitemShut {NoStop}%
\bibitem [{RSM()}]{RSM}%
  \BibitemOpen
  \href@noop {} {}\bibinfo {note} {By {XRD} reciprocal space mapping on three
  representative samples ($\text{La}/\text{Al} = 0.908, \, 0.996, \, 1.155$),
  we confirmed that the {LAO} films were fully strained to the STO (001)
  substrate}\BibitemShut {NoStop}%
\bibitem [{\citenamefont {Luo}, \citenamefont {Wang},\ and\ \citenamefont
  {Zheng}(2009)}]{PhysRevB-80-104115}%
  \BibitemOpen
  \bibfield  {author} {\bibinfo {author} {\bibfnamefont {X.}~\bibnamefont
  {Luo}}, \bibinfo {author} {\bibfnamefont {B.}~\bibnamefont {Wang}}, \ and\
  \bibinfo {author} {\bibfnamefont {Y.}~\bibnamefont {Zheng}},\ }\href
  {\doibase 10.1103/PhysRevB.80.104115} {\bibfield  {journal} {\bibinfo
  {journal} {Phys. Rev. B}\ }\textbf {\bibinfo {volume} {80}},\ \bibinfo
  {pages} {104115} (\bibinfo {year} {2009})}\BibitemShut {NoStop}%
\bibitem [{\citenamefont {Kr{\"{o}}ger}\ and\ \citenamefont
  {Vink}(1956)}]{SolidStatePhys-3-307}%
  \BibitemOpen
  \bibfield  {author} {\bibinfo {author} {\bibfnamefont {F.~A.}\ \bibnamefont
  {Kr{\"{o}}ger}}\ and\ \bibinfo {author} {\bibfnamefont {H.~J.}\ \bibnamefont
  {Vink}},\ }\href@noop {} {\bibfield  {journal} {\bibinfo  {journal} {Solid
  State Phys.}\ }\textbf {\bibinfo {volume} {3}},\ \bibinfo {pages} {307}
  (\bibinfo {year} {1956})}\BibitemShut {NoStop}%
\bibitem [{\citenamefont {Reinle-Schmitt}\ \emph {et~al.}(2012)\citenamefont
  {Reinle-Schmitt}, \citenamefont {Cancellieri}, \citenamefont {Li},
  \citenamefont {Fontaine}, \citenamefont {Medarde}, \citenamefont
  {Pomjakushina}, \citenamefont {Schneider}, \citenamefont {Gariglio},
  \citenamefont {Ghosez}, \citenamefont {Triscone},\ and\ \citenamefont
  {Willmott}}]{arXiv:1112.3532}%
  \BibitemOpen
  \bibfield  {author} {\bibinfo {author} {\bibfnamefont {M.~L.}\ \bibnamefont
  {Reinle-Schmitt}}, \bibinfo {author} {\bibfnamefont {C.}~\bibnamefont
  {Cancellieri}}, \bibinfo {author} {\bibfnamefont {D.}~\bibnamefont {Li}},
  \bibinfo {author} {\bibfnamefont {D.}~\bibnamefont {Fontaine}}, \bibinfo
  {author} {\bibfnamefont {M.}~\bibnamefont {Medarde}}, \bibinfo {author}
  {\bibfnamefont {E.}~\bibnamefont {Pomjakushina}}, \bibinfo {author}
  {\bibfnamefont {C.~W.}\ \bibnamefont {Schneider}}, \bibinfo {author}
  {\bibfnamefont {S.}~\bibnamefont {Gariglio}}, \bibinfo {author}
  {\bibfnamefont {P.}~\bibnamefont {Ghosez}}, \bibinfo {author} {\bibfnamefont
  {J.-M.}\ \bibnamefont {Triscone}}, \ and\ \bibinfo {author} {\bibfnamefont
  {P.~R.}\ \bibnamefont {Willmott}},\ }\href@noop {} {\bibfield  {journal}
  {\bibinfo  {journal} {Nature Commun.}\ }\textbf {\bibinfo {volume} {3}},\
  \bibinfo {pages} {932} (\bibinfo {year} {2012})}\BibitemShut {NoStop}%
\bibitem [{\citenamefont {Harrison}\ \emph {et~al.}(1978)\citenamefont
  {Harrison}, \citenamefont {Kraut}, \citenamefont {Waldrop},\ and\
  \citenamefont {Grant}}]{PhysRevB-18-4402}%
  \BibitemOpen
  \bibfield  {author} {\bibinfo {author} {\bibfnamefont {W.~A.}\ \bibnamefont
  {Harrison}}, \bibinfo {author} {\bibfnamefont {E.~A.}\ \bibnamefont {Kraut}},
  \bibinfo {author} {\bibfnamefont {J.~R.}\ \bibnamefont {Waldrop}}, \ and\
  \bibinfo {author} {\bibfnamefont {R.~W.}\ \bibnamefont {Grant}},\ }\href
  {\doibase 10.1103/PhysRevB.18.4402} {\bibfield  {journal} {\bibinfo
  {journal} {Phys. Rev. B}\ }\textbf {\bibinfo {volume} {18}},\ \bibinfo
  {pages} {4402} (\bibinfo {year} {1978})}\BibitemShut {NoStop}%
\end{thebibliography}
%

\end{document}